\documentstyle[epsf]{l-aa}

\begin{document}

\thesaurus{08.16.7} 
   
\title{Radius \& Distance Estimates of the Isolated Neutron Stars Geminga \& PSR B0656+14 using Optical Photometry}

\author{A. Golden \inst{1}  A. Shearer \inst{2} }

\offprints{A. Golden, agolden@physics.ucg.ie}

\institute{ Department of Physics, National University of Ireland, Galway, Ireland
	    \and Information Technology Centre, National University of Ireland, Galway, Ireland}

\date{Received ...; Accepted ...}

\maketitle
\markboth{Golden, A. \& Shearer, A.: {\it R/d} Estimates for Geminga \& PSR B0656+14 using Optical Photometry}{}
   
%________________________________________________________________

\begin{abstract}

Integrated ground-based and HST optical studies of isolated neutron stars have
provided important independent datasets in the determination of emission activity, 
particularly in the fitting of anticipated Rayleigh-Jeans extrapolations from 
EUV/soft X-ray datasets, despite their intrinsic faintness. Differentiation of the pulsed and
unpulsed fluxes and consequently of the nonthermal and thermal modes of emission
could provide definitive data with which to constrain this blackbody continuum. Based upon
high speed photometric observations of Geminga and PSR B0656+14 in the $B$ band, 
we have combined upper limits of unpulsed emission with recently published model-fits with
a view to assessing possible implications for the $R/d$ parameter. For Geminga, with
a known distance of $\sim$ 160 pc, we find that $R_{\infty}$ $\leq$ 9.5 km for a $\sim$ 
blackbody source, and $R_{\infty}$ $\leq$ 10.0 km with the presence of a magnetized $H$ 
atmosphere. In addition, we suggest that PSR B0656+14 is some $\sim$ 4 - 5 times closer 
than the 760 pc estimated from DM measurements alone.

\end{abstract}

\section{Introduction}
 
Uncertainty regarding the behaviour of nuclear matter in the
deep neutron star interior has compromised
a complete description of the dense matter equation of state (EOS).
Various theoretical models in circulation predict a range of macroscopic
observables, such as masses, radii, temperatures and maximum rotation rates, 
and as such are open to scrutiny with empirical data. Relativistic
effects complicate such observations, with
the true radius, $R$, related to the {\it apparent} radius at a distance 
( d $\rightarrow \infty$ ) as
\begin{equation}
R_{\infty} = (1+z)R = R / \sqrt{1 - 2GM/Rc^{2}},
\end{equation}
with $M$ being the mass of the compact object, and $z$ the associated gravitational redshift. 
Theory to date indicates that 7km $<$ $R_{\infty}$ $<$ 20km, dependent on the stiffness of the EOS
(e.g. \cite{lind92}). Binary studies suggest a common value of neutron star mass 
approximately that of the canonical estimate of 1.4 $M_{\odot}$ (\cite{vank95}).
Typically, the model neutron star is assumed to have $M$ $\sim$ 1.4 $M_{\odot}$
and $R_{\infty}$ $\sim$ 13km ($R$ $\sim$ 10 km). Based on observations of the neutron 
star's thermal emission in the extreme UV (EUV) and soft X-ray bands, 
it should in principle be feasible to compute the ideal blackbody spectral energy 
distribution (SED) as a function of ($T_{\rm surface}$, $N_{\rm H}$ \& $R/d$), and 
for a known distance $d$, an estimate of the apparent neutron star radius.
It is generally agreed that such a measurement would have a profound impact
in constraining EOS models. However, strong galactic $HI$ absorption at these 
wavelengths restricts observations to the closest neutron stars, and uncertainty 
in X-ray detector sensitivities at these low ($\leq$ 0.2 keV) energies 
( e.g. \cite{wal98}) has compromised attempts to accurately determine estimates 
for $N_{\rm H}$ and $T_{\rm surface}$.
Furthermore, such an analysis may be complicated by neutron star phenomenology, such as
atmospheric opacity effects, an active magnetosphere, hot polar cap regions and 
accretion processes. It is critical to separate the various contributions, so as 
to estimate the genuine total surface component. For the older, isolated neutron 
stars (INS), phase-resolved studies in the soft X-ray, EUV and optical wavebands 
can provide such a SED, and thus a real possibility of determining $R_{\infty}$. 
The Rayleigh-Jeans tail in the optical regime provides stringent constraints to any
SED model-fit, and in this waveband atmospheric opacity effects are expected to have the most 
noticeable impact. In the X-ray regime where the blackbody continuum peaks, low $Z$ atmospheres
preferentially transmit radiation from the lower, hotter regions of the
photosphere, producing a X-ray spectrum suggestive of a 'hotter' source (\cite{rom87}). 
The Rayleigh-Jeans tail is unaffected by this deviation, and discrepancies between optical 
spectrophotometry and higher energy extrapolations may be rigorously tested (\cite{pav98}).
However the intrinsic faintness of these astrophysical objects in the optical regime coupled 
with the limitations of current technology have restricted previous optical studies to 
deep integrated photometry.  These observations have in some ways aided such
thermal continuum studies (e.g. \cite{wal97}), but the differentiation of 
pulsed (predominantly nonthermal) and unpulsed (thermal) components would be ideal in 
a more rigorous treatment. 
Recently, the TRIFFID high speed optical 
photometer detected pulsations in the $B$ band from the optical counterparts of 
the middle aged pulsars Geminga and PSR B0656+14 (\cite{shear97}, \cite{shear98}). 
Both light curves are highly pulsed and suggest a dominant 
nonthermal mode of emission optically. Despite their extreme 
faintness, it was possible in both cases to determine upper limits to each
pulsar's thermal component of emission. In this
letter, we combine these unpulsed limits with the results of recently published
ground-based and HST photometric analysis on these two pulsars (\cite{pav97}, 
\cite{mart98} hereafter $PWC97$ \& $MHS98$), and derive radius/distance
estimates for both. In particular, for the case of Geminga with a known parallax
distance of $\sim$ 160 pc, we provide for the first time an upper limit for 
$R_{\infty}$ based upon such phase-resolved photometry.

We approach the analysis of the optical photometry by
assuming a model fit incorporating both nonthermal power-law and thermal components of
emission in the UBVRI regime, as originally adopted by $PWC97$. This
two-component model fit is defined as:
\begin{equation}
f(\nu)=\left[f_0 \left(\frac{\nu}{\nu_0}\right)^{-\alpha} + g_0
\left(\frac{\nu}{\nu_0}\right)^2\right]\times 10^{-0.4 A(\nu)}~,
\end{equation}
with $\nu_0$ = 8.766 $\times 10^{14}$ Hz an arbitrary reference frequency, and $A(\nu)$ the interstellar 
extinction determined following Savage \& Mathis (1979). 
$g_0$ and $f_0$ are taken to be the values of the nonthermal and thermal fluxes at $\nu=\nu_0$. 
The latter can be expressed, for the chosen value of $\nu_0$, as
\begin{equation}
g_0=
3.116\times10^{-31} G ~\frac{{\rm erg}}{{\rm cm}^{2}~{\rm s}~{\rm Hz}}~,
\qquad G \equiv T_6 \left(\frac{R_{10}}{d_{500}}\right)^2
\end{equation}
where $T=10^6 T_6$~K is the apparent neutron star brightness temperature,
$R_\infty = 10 R_{10}$~km the radius and $d=500 d_{500}$~pc the distance to the 
blackbody. 
Model fits for both pulsars have suggested the presence of both emission modes, 
yet uncertainty in the optimum ($T_{\rm surface}$,$N_{\rm H}$, $R/d$) solution
based on EUV/soft X-ray datasets restricts accurate differentiation. By providing 
upper bounds for the unpulsed flux, and indirectly $N_{\rm H}$ via the estimated $A(\nu)$ 
towards the pulsar, we can derive an {\it independent} estimate of $G$
and in this way, constrain ($T_{\rm surface}$, $R/d$) space. The use of the lower 
bounds to $T_{\rm surface}$ for a given neutron star would then yield {\it upper} 
bounds to the $R/d$ parameter.

\section{Geminga}

PSR 0633+17, or Geminga, provides one of the most ideal neutron stars for such an
analysis, as HST observations have determined its distance by parallax to 
$159^{+59}_{-34}$ pc (\cite{car96}). However, it is by no means clear precisely
what the nature of its emission processes are, despite considerable
observational efforts from the optical to $\gamma$-rays. Original ROSAT
observations suggested dominant thermal emission from the surface modulated
by a hotter polar cap (\cite{hal93}), although later observations especially those of {\it ASCA}
indicated the emission was thermal and nonthermal in origin (\cite{hal97}). Disagreement remains on the optimum SED fits to
data from ROSAT, EUVE and ASCA with the pulsar's surface 
temperature within $( 2.5 - 6 )\times10^{5}$ K. There have been further suggestions
that the pulsar's thermal emission is strongly affected by magnetospheric cyclotron
resonance blanketing (\cite{wang98}), which would severely compromise any estimate
of the {\it true} neutron star surface temperature. Optical observations suggest an
unusual SED, with deviations in the expected functional
form, possibly as a result of ion cyclotron absorption/emission processes.
The most recent of such observations ($MHS98$), made using the Keck LRIS, spanned 370-800 nm
and yielded a flat power law shape ($f_{\nu}$ $\propto$ ~$\nu^{-0.8}$) and a noticeable broad dip at
630-650 nm with perhaps a slight modulation at $V$, $B$ $\&$ $I$ as advanced by Bignami et al. (1996).
The composite power-law may be fitted by either a combined
blackbody and power-law (i.e. nonthermal emission with $\alpha$ $\sim$ $1.9 \pm 0.6$) or a blackbody plus
global ion cyclotron emission. The former model is undoubtedly the
most likely, following the discovery of a highly pulsed lightcurve in the $B$ band 
(\cite{shear98}). Fig. 1. shows the model-fit spectrum of $MHS98$ with 
the results of Shearer et al. (1998).
   
\begin{figure}
\epsfysize 3.0truein
\epsffile{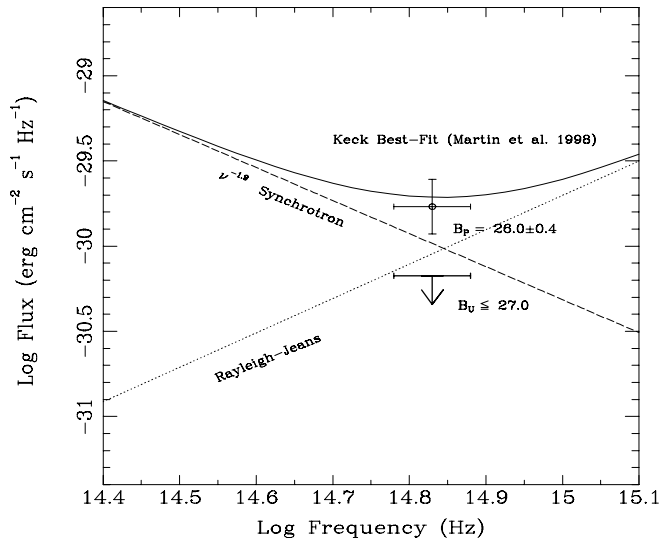}
\caption{Keck LRIS Optical spectrum of Geminga, with best-fit two component model of Martin et al. (1998) and
TRIFFID/BTA pulsed and unpulsed flux estimates in $B$ (Shearer et al. 1998)}
\label{vband}
\end{figure} 

A value of $G$ = 4.4 is obtained by applying a solution in the form of (3) to the best-fit
Keck observations of $MHS98$, setting A($\nu$) $\sim$ 0 due to the pulsar's close proximity.
Introducing our independently estimated unpulsed upper limit in the
$B$ band of $6.77\times 10^{-31} ~{\rm ergs} ~{\rm cm^{2}} ~{\rm s^{-1}} ~{\rm Hz^{-1}}$, we conclude
with $G$ $\leq$ $3.6 \pm 0.1$, allowing errors in $G$ for the spectral response
of the MAMA/$B$-filter combination.
The decrease can be inferred as either a reduction in the emission area
or a drop in $T_{\rm surface}$ ($d$ $\sim$ 160 pc). As a consequence, one expects
the nonthermal component to increase proportionately in the original model fit, if this
unpulsed estimate is representative of a general surface based thermal emission.
If we assume that $T_{\rm surface}$ remains unchanged, then the obvious conclusion is that $R$ has been overestimated.
In fact, using $G$ $\leq$ 3.6, $T_{\rm surface}$ $\sim$ $4.0\times 10^{5}$ K (MHS98) and $d$ = $159^{+59}_{-34}$ pc
suggests $R_{\infty}$ $\leq$ $9.5 ^{+3.5}_{-2.0}$ km for a $\sim$ blackbody source (indistinguishable from a Fe/Si
atmosphere - see \cite{pav98}).
An alternative explanation for the change in $G$ could be understood under the
assumption of atmospheric opacity effects for the total surface thermal emission
at X-ray energies. As $PWC97$ point out, there is no ideal way to
reconcile the actual brightness temperature with the estimated X-ray $T_{\rm surface}$. 
Fits with magnetized ( $\sim$ $10^{12}$ G ) $H$ atmospheres suggest
that a spectrum in the optical regime can be fitted as a Rayleigh-Jeans
tail with $T_{\rm surface} = 0.9 T_{\rm eff}$ ($PWC97$). Applying this correction yields 
$R_{\infty}$ $\leq$ $10.0 ^{+3.8}_{-2.1}$ km.

\section{PSR B0656+14}

An extensive literature exists devoted to the analysis of EXOSAT,
ROSAT and EUVE observations of the pulsar PSR B0656+14. The consensus is of
thermal emission from the surface, modulated by emission from a hot polar cap
and from nonthermal activity in the magnetosphere. However, its distance estimate is by no
means exact ranging from $\sim$ 100 to 760 pc (\cite{car94}) - the latter based upon a
DM fit to the uncertain galactic electron model in this vicinity (Taylor et al. 1993). 
Applying a common distance estimate of $\sim$ 500 pc, a range of 
solutions are possible in ($T_{\rm surface}$, $N_{\rm H}$) space assuming the canonical model, and this is reflected in the 
literature, with $T_{\rm surface}$ = (3 - 9)$\times 10^{5}$ K and $N_{\rm H}$ = (0.5 - 2.0)$\times 10^{20} 
cm^{2}$ (\cite{fin92}, \cite{and93}, \cite{gre96}). Early limited observations using the $NTT$ 
(\cite{car94}) and the $HST$ (\cite{pav96}) indicated that the optical counterpart's emission 
was predominately nonthermal in nature, and a subsequent two-component model fit 
to detailed $HST$ and ground-based photometry spanning the UBVRI regime by $PWC97$ 
substantiated this assessment - although the detection of highly pulsed emission
in the $B$ band (\cite{shear97}) from the optical counterpart unequivocally confirmed 
this hypothesis. The upper limit on the unpulsed flux from the resulting lightcurve 
was estimated to be $8\times 10^{-31} ~{\rm ergs} ~{\rm cm^{-2}}~ {\rm s^{-1}}~ {\rm Hz^{-1}}$. Fig. 2. shows the 
$PWC97$ model fit with these pulsed/unpulsed fluxes.

$PWC97$ fitted the observed UBVRI spectrum with a two-component model following
the formalism of (2) and (3). Applying the interstellar extinctions determined
for the three estimated colour excesses towards the pulsar of $E(B - V)$ = 0.01, 
0.03 and 0.05 yielded best fit values of $G$ = 3.0, 3.7 and 4.3  respectively. 
Taking $g_{0}$ = $ 8\times 10^{-31} ~{\rm ergs} ~{\rm cm^{-2}}~ {\rm s^{-1}}~ {\rm Hz^{-1}}$, (2) may be 
rearranged in terms of $G$, and solved for the three $E(B - V)$ estimates, concluding 
with $G$ = 4.4, 4.8 and 5.2 (all $\pm 0.1$).
respectively. An increase of this $G$ parameter is consistent with either an 
increase in the expected $T_{\rm surface}$, emission area or a decrease in the pulsar's 
distance.

\begin{figure}
\epsfysize 3.0truein
\epsffile{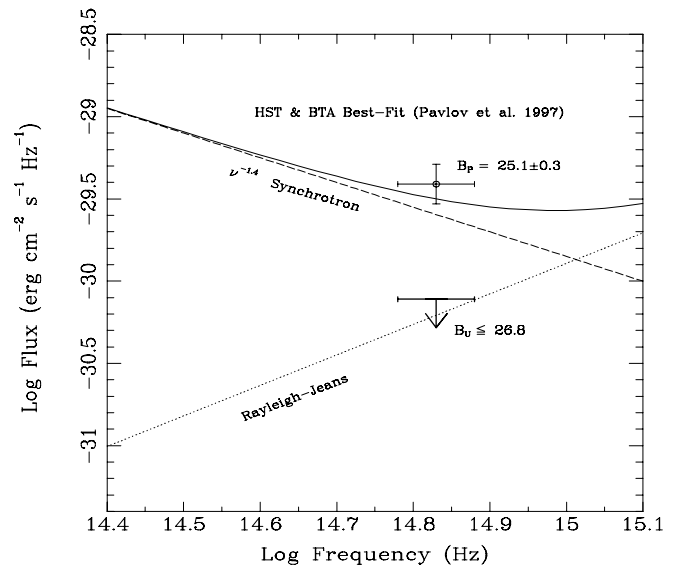}

\caption{HST \& 6m BTA based photometry and best-fit two component model of PSR B0656+14 for the
colour excess $E(B-V)$ = 0.03 (Pavlov et al. 1997), with TRIFFID/BTA pulsed and unpulsed flux estimates in $B$
superimposed (Shearer et al. 1997)}
\label{vband}
\end{figure}

In terms of previous model fits, $G$ ranges from 0.9 (\cite{fin92}), 2.1 (\cite{gre96}) to 2.6 
(\cite{and93}), the latter incorporating a magnetised ($\sim 10^{12}$ G) $H$ atmosphere. As 
observations have accrued, and uncertainties in X-ray detector sensitivities have been addressed, 
the trend has been a decrease in the derived surface temperature, from $\sim 9.0\times 
10^{5}$K to $\sim 5.0\times 10^{5}$K. Recent work by Edelstein et al. (1998), which substituted 
EUVE DS data in place of the uncertain low energy ROSAT PSPC channels, has yielded 
a ($T_{\rm surface}$,$N_{\rm H}$) space differing markedly from earlier ROSAT results alone. 
Combining this new parameter space with independent model fits incorporating
the $B$ unpulsed upper limit suggests $T_{\rm surface}$ $\geq$ 5.0$\times 10^{5}$ K (\cite{gol98}).
It is possible to constrain $R/d$ for B0656+14 in two ways - firstly by determining apparent
expected radii using both column density and $DM$ derived distances, and secondly by 
applying canonical or otherwise derived radius limits to yield optimum distance scales. 
$N_{\rm H}$ estimates towards PSR B0656+14, although by no means certain, suggest that the pulsar is 
$\sim$ 250-280 pc (\cite{and93}, \cite{edel98}), rather closer than the DM derived distance of 
$760 \pm 190$ pc. By manipulation of (3) with $T_{\rm surface}$ $\geq$ 5.0$\times 10^{5}$ K and the
range of $G$ parameters determined from the unpulsed upper limit, the expected radial estimates 
for the $N_{\rm H}$ distances are 14.7 $<$ $R_{\infty}$ $<$ 17.7 km, and substantially greater 
in the case of the radio derived distance. We note that such radii estimates are {\it in excess} of the 
14 km upper limit determined by Walter et al. (1997) for the old INS RXJ185635-3754.
Alternatively, applying the ideal canonical $R_{\infty}$ $\sim$ 13km, manipulation of (3)
as above suggests that 205 $\leq$ $d$ $\leq$ 227 pc based upon the range of $E(B-V)$ estimates.
This supports the conclusions of $PWC97$ although placing the pulsar somewhat in closer proximity than had been 
originally thought. Indeed if one was to consider the proposed estimate of $R_{\infty}$ $\sim$ $9.5 ^{+3.5}_{-2.0}$ 
km for Geminga as a working upper limit, this would place PSR B0656+14 at a distance of 
no less than $d$ = $152 ^{+55}_{-32}$ for the suggested optimum colour index of $E(B-V)$ = 0.03 (PWC97)

\section{Conclusions}

The acquisition and analysis of optical photometric data on INS
has been shown to provide an independent dataset from which constraints to the optimum
thermal spectral energy distribution may be applied. We have attempted for the
first time to apply the phase-resolved optical flux consistent with that expected {\it specifically}
from the thermal component to the previous integrated analysis of the pulsars Geminga
and PSR B0656+14. Despite being upper limits to the unpulsed optical flux in the $B$ band,
we find that in both cases, the resulting blackbody spectral distribution is constrained
to the extent that we can set upper limits to the $R/d$ parameter for both
neutron stars. For the case of Geminga, with a known parallax derived distance of $\sim$ 160 pc and
using the lower $T_{\rm surface}$ limit, we suggest that $R_{\infty}$ $\leq$ 9.5 km for a $\sim$ blackbody source,
and $R_{\infty}$ $\leq$ 10.0 km with the presence of a magnetized $H$ atmosphere. Previous
work using these unpulsed upper limits has suggested a $T_{\rm surface}$ $\geq$ 5.0$\times 10^{5}$ K for the pulsar B0656+14
(Golden, 1998), and under the assumption of $R_{\infty}$ $\sim$ 13km, places the pulsar at $\sim$ 210 pc - in contrast to
the $DM$ derived distance of 760 pc. Assuming the neutron star has $R_{\infty}$ $\leq$ 9.5 km, then
this limits the distance to $\sim$ 160 pc. This suggests the possibility that the pulsar may
be a viable candidate for a parallax measurement attempt using the HST. 

Despite using upper limits to the unpulsed fluxes of these two INS, 
we have been successful in indicating how important such definitive measurements are
in the rigorous derivation of a given neutron star's thermal parameters. In this
way, we have independently provided limits to the radius of one, Geminga, and
the distance of another, PSR B0656+14 and shown the promise to future studies
this observational technique offers. This short report documenting these results
is most timely, following on from a recent forum on the study of INS
in this energy regime, at which the future importance of such high speed
photometric studies was stressed (\cite{rom98}), and where a recent theoretical analysis based upon
the discovery of RXJ185635-3754 has shown that neutron star radii should be ideally $\leq$ 10 km
if one can hope to effectively constrain the EOS (\cite{an98}). That our results
point to such a radius limit only emphasises the need for further
high speed photometry of both pulsars, particularly Geminga, so as to provide
definitive, rather than upper limit, fluxes of their unpulsed emission.

%__________________________________________________________________

\begin{acknowledgements}
The authors acknowledge the financial support of {\it Enterprise Ireland}.
The preliminary work was done by AG whilst a visiting RA at UC Berkeley
and Stu Bowyer is thanked for his support during this time.

\end{acknowledgements}

\end{document}